\documentclass[pra,twocolumn,superscriptaddress,floatfix,longbibliography,nofootinbib]{revtex4-1}

\usepackage[utf8]{inputenc}
\usepackage[english]{babel}
\usepackage{amsmath,amssymb,mathtools,bm}
\usepackage{graphicx}
\usepackage[dvipsnames]{xcolor}
\usepackage{physics}
\usepackage{booktabs}
\usepackage{microtype}
\usepackage{mathpazo}
\usepackage[colorlinks=true,citecolor=Blue,urlcolor=Blue,linkcolor=Blue]{hyperref}

\newif\ifshowchanges
\showchangesfalse
\definecolor{changecolor}{RGB}{200,0,0}
\DeclareRobustCommand{\chg}[1]{\ifshowchanges{\color{changecolor}#1}\else#1\fi}

\newcommand{\ii}{\mathrm{i}}
\newcommand{\bhat}[1]{\hat{#1}}

\newcommand{\Hloc}{\bhat{V}_{\mathrm{loc}}}
\newcommand{\Jtot}{\mathcal{J}}
\newcommand{\JR}{\mathcal{J}_{R}}
\newcommand{\JL}{\mathcal{J}_{L}}
\newcommand{\Jinc}{\mathcal{J}_{0}}
\newcommand{\Dr}{\Delta_{r}}
\newcommand{\td}{t_{d}}
\newcommand{\kr}{k_{r}}
\newcommand{\vg}{v_{g}}
\newcommand{\hc}{\mathrm{H.c.}}

\usepackage{tikz}
\usetikzlibrary{arrows.meta,decorations.pathmorphing,shadows}

\begin{document}
\raggedbottom

\title{Ground-State Dressing and Delayed Feedback\texorpdfstring{\\}{ }of a Chiral Giant Atom in Ultrastrong Waveguide QED}

\author{Giuseppe Musolino}
\affiliation{Dipartimento di Scienze Matematiche e Informatiche, Scienze Fisiche e Scienze della Terra, Universit\`{a} di Messina, I-98166 Messina, Italy}

\author{Daniele Lamberto}
\affiliation{Quantum Information Physics Theory Research Team, RIKEN Center for Quantum Computing, Wakoshi, Saitama, 351-0198, Japan}

\author{Salvatore Savasta}
\affiliation{Dipartimento di Scienze Matematiche e Informatiche, Scienze Fisiche e Scienze della Terra, Universit\`{a} di Messina, I-98166 Messina, Italy}

\author{Federico Roccati}
\affiliation{Quantum Theory Group, Dipartimento di Fisica e Chimica -- Emilio Segr\`e,
	Universit\`a degli Studi di Palermo, I-90123 Palermo, Italy}

\begin{abstract}
We study a chiral giant atom ultrastrongly coupled to a
one-dimensional waveguide at two spatially separated points.
At a relative coupling phase $\phi=\pi/2$, the total spectral
density is independent of the distance between the coupling
points, whereas its directional components are not.
Using a variational polaron transformation, we show that
the renormalized transition frequency and the total
ground-state dressing inherit this independence, while
the virtual-photon cloud remains chiral.
A sudden switch-off of the coupling releases this cloud
as directional radiation.
We also derive the spontaneous-emission dynamics in the
polaron frame, showing that, after neglecting the local
photonic potential, the memory kernel becomes distance
independent at the same phase.
\end{abstract}

\maketitle

\section{Introduction}

Giant atoms couple to a waveguide at multiple spatially
separated points, giving rise to interference effects that
depend on the distance between the coupling points
\cite{Kockum2021,Kockum2014,Kockum2018,Guo2017,Andersson2019,Kannan2020,Leonforte_2025,Soro,Du,Bonsel2026fibonacciwaveguide,Almanakly2026}.
In particular, the propagation time between the coupling
points can produce delayed self-interaction and
nonexponential spontaneous emission~\cite{Cillufo,Scheremet,Lin,Guo,guo2,Facchi,Zhao1,Cheng,Gu7,Cilluffo2026,Diekmann2026}.
When the local couplings carry a relative complex phase,
the interaction becomes chiral, with different coupling
strengths to the two propagation directions
\cite{Ramos2016,wang,Wang_2022,Lodahl2017,SoroKockum2022,JoshiMirhosseini2023,Chen2022,PhysRevA.109.063703,PhysRevA.107.063703,Du_2023}.
In the weak-coupling regime, this phase can also eliminate
the delayed self-interaction, even for a finite propagation
time between the coupling points \cite{Roccati2024,Chang2025}.

The ultrastrong-coupling (USC) regime requires going
beyond the rotating-wave approximation (RWA), since
counterrotating processes renormalize the atomic transition
frequency and dress the ground state with virtual photons
\cite{FriskKockum2019}.
Variational polaron transformations have been employed
to describe equilibrium properties and low-energy dynamics
in this regime
\cite{DiazCamacho2016,SanchezBurillo2019,RomanRoche2020,
GonzalezGutierrez2021,Terradas2022}.
In particular, Ref.~\cite{Terradas2022} developed a polaron
description of nonchiral giant atoms in USC.
Spontaneous-emission spectra beyond the rotating-wave
approximation have also been investigated
\cite{YanLu2024}, while recent work has considered
non-Markovian giant-atom dynamics beyond this approximation
\cite{YuStrunzNimmrichter2026} and the role of spectral
bound states in strongly coupled giant-atom systems
\cite{WuAn2026}.
The combined effects of chiral coupling phases and
ultrastrong ground-state dressing remain less explored.

In this work, we consider a two-level emitter ultrastrongly coupled
to a waveguide at two points with equal coupling strengths
and a relative phase $\phi$.
We show that the total spectral density depends on this
phase through $\cos\phi$, whereas the difference between
the directional spectral densities is proportional to
$\sin\phi$.
Consequently, at $\phi=\pi/2$, the total
spectral density becomes independent of the distance
between the coupling points.
Within a variational polaron description, the renormalized
transition frequency, the ground-state atomic population,
and the total number of virtual photons inherit this
distance independence.
Nevertheless, the momentum distribution of the virtual
photons remains chiral.
We show how a sudden switch-off of the light--matter
coupling converts this asymmetry into directional
propagating radiation, extending the use of coupling
quenches to the present setting \cite{SanchezBurillo2019}.

We also investigate spontaneous emission using the
number-conserving low-energy polaron Hamiltonian.
After neglecting the local photonic potential generated
by the transformation, we derive a memory equation
whose kernel becomes independent of the distance
between the coupling points at $\phi=\pi/2$.
In the broad-band, linear-dispersion limit, this reduces
to a delay equation without delayed self-interaction,
as in the weak-coupling regime, but with renormalized
parameters.
For a finite-band waveguide, reservoir memory can
still produce nonexponential dynamics.

The paper is organized as follows.
Sections~\ref{sec:model} and \ref{sec:polaron} introduce
the model and the polaron description.
Section~\ref{sec:equilibrium} addresses equilibrium
properties, Sec.~\ref{sec:chirality} discusses the
directional virtual-photon cloud and its release after
a quench, and Sec.~\ref{sec:decay} examines spontaneous
emission and delayed feedback.
Additional derivations and the bound-state analysis
are presented in the Appendixes.

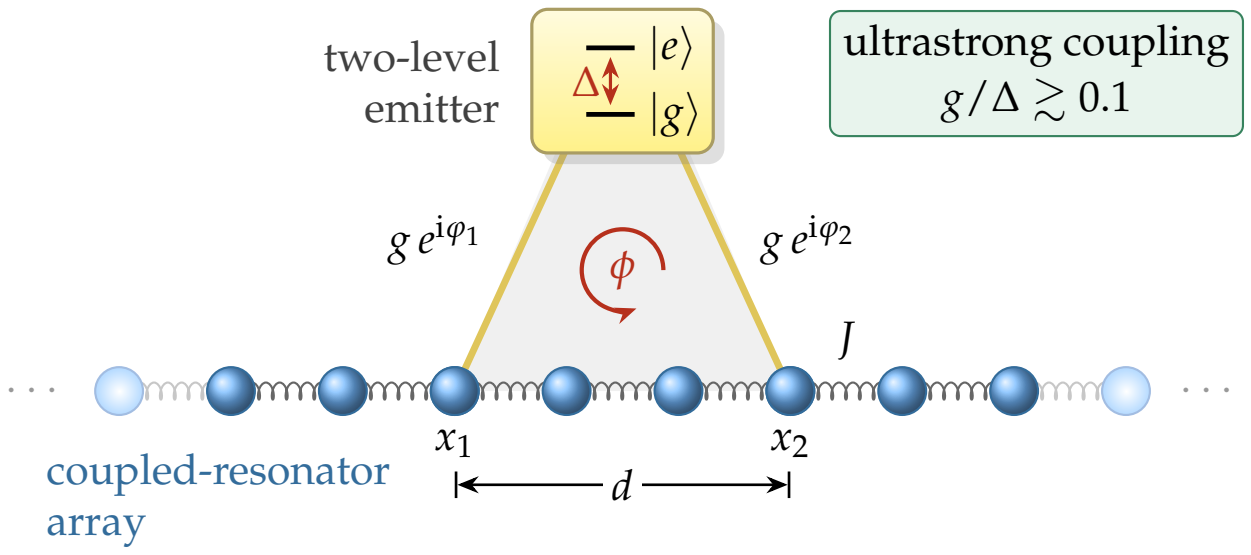
\begin{figure}[t]
    \centering
    \begin{tikzpicture}[
        >={Stealth[length=4pt,width=3.2pt]},
        font=\footnotesize,
        res/.style={circle,inner sep=0pt,minimum size=9pt,
                    ball color=RoyalBlue!55,draw=RoyalBlue!70!black,line width=0.4pt},
        bond/.style={decorate,decoration={coil,aspect=0.45,segment length=2.6pt,amplitude=1.6pt},
                     line width=0.4pt,draw=black!60},
        leg/.style={line width=1.3pt,draw=Goldenrod!85!black,line cap=round},
        photon/.style={decorate,decoration={snake,amplitude=1.5pt,segment length=6pt,post length=4pt},
                       ->,line width=0.7pt},
      ]
      \def\s{0.74}            
      \def\ya{2.05}           
      \coordinate (A) at (0,\ya);
      \coordinate (x1) at (-1.5*\s,0);
      \coordinate (x2) at ( 1.5*\s,0);

      \fill[black!6,rounded corners=1pt] (x1) -- (-0.42,\ya-0.42) -- (0.42,\ya-0.42) -- (x2) -- cycle;
      \draw[->,line width=0.8pt,BrickRed] (0.27,0.80) arc[start angle=0,end angle=290,radius=0.27];
      \node[BrickRed] at (0,0.80) {$\phi$};

      \foreach \i in {-5,...,3}{
        \pgfmathsetmacro{\fade}{ifthenelse(\i==-5 || \i==3,1,0)}
        \ifnum\fade=1 \def\bc{black!22}\else\def\bc{black!60}\fi
        \draw[bond,draw=\bc] ({(\i+0.5)*\s},0) -- ({(\i+1.5)*\s},0);
      }
      \node[black!40] at (-5.25*\s,0) {$\cdots$};
      \node[black!40] at ( 5.25*\s,0) {$\cdots$};
      \foreach \i in {-5,...,4}{
        \pgfmathtruncatemacro{\fade}{ifthenelse(\i==-5 || \i==4,1,0)}
        \ifnum\fade=1
          \shade[inner color=white,outer color=RoyalBlue!30,draw=RoyalBlue!35,line width=0.4pt] ({(\i+0.5)*\s},0) circle (4.5pt);
        \else
          \node[res] at ({(\i+0.5)*\s},0) {};
        \fi
      }
      \node[above=1pt] at (2.0*\s,0.06) {$J$};

      \draw[leg] (x1) -- (-0.36,\ya-0.42);
      \draw[leg] (x2) -- ( 0.36,\ya-0.42);
      \node[res] at (x1) {};
      \node[res] at (x2) {};
      \node[anchor=east] at (-0.80,1.0) {$g\,e^{\mathrm{i}\varphi_1}$};
      \node[anchor=west] at ( 0.80,1.0) {$g\,e^{\mathrm{i}\varphi_2}$};

      \node[draw=Goldenrod!60!black,line width=0.6pt,rounded corners=3pt,
            top color=Goldenrod!15,bottom color=Goldenrod!55,
            minimum width=1.2cm,minimum height=0.95cm,drop shadow={opacity=0.25}] (box) at (A) {};
      \draw[line width=0.9pt] (-0.24,\ya+0.22) -- (0.08,\ya+0.22) node[right=-1pt] {$\ket{e}$};
      \draw[line width=0.9pt] (-0.24,\ya-0.22) -- (0.08,\ya-0.22) node[right=-1pt] {$\ket{g}$};
      \draw[<->,line width=0.6pt,BrickRed] (-0.08,\ya-0.16) -- (-0.08,\ya+0.16);
      \node[BrickRed,anchor=east,inner sep=1pt] at (-0.10,\ya) {$\Delta$};

      \node[below=4pt] at (x1) {$x_1$};
      \node[below=4pt] at (x2) {$x_2$};
      \draw[|<->|,line width=0.5pt]
         ([yshift=-0.62cm]x1) -- node[fill=white,inner sep=1.5pt] {$d$} ([yshift=-0.62cm]x2);

      \node[anchor=east,align=right,text=black!75] at (-0.70,\ya) {two-level\\emitter};
      \node[anchor=north west,align=left,text=RoyalBlue!70!black] at (-5.3*\s,-0.30) {coupled-resonator\\array};
      \node[draw=ForestGreen!70!black,rounded corners=2pt,fill=ForestGreen!8,
            align=center,inner sep=2.5pt] at (3.7*\s,\ya+0.05)
            {ultrastrong coupling\\[1pt] $g/\Delta\gtrsim0.1$};
    \end{tikzpicture}
    \caption{Chiral giant atom coupled at two points to a waveguide, implemented as an array of coupled resonators. The two local interactions have equal magnitude $g$ and relative phase $\phi=\varphi_2-\varphi_1$, while $d$ is the distance between the coupling points.}
    \label{fig:setup}
\end{figure}

\section{Model and  spectral densities}
\label{sec:model}

We consider a single two-level emitter coupled to a one dimensional bosonic waveguide at two distinct points, a configuration known as a `giant' emitter~\cite{Vadiraj}. We model the waveguide as a coupled-cavity array~\cite{Jouanny2025}, i.e.\ a chain of $N$ identical resonators, and the emitter as being coupled to the cavities at sites $x_1$ and $x_2$. In the site (position) basis, the total Hamiltonian reads (setting $\hbar=1$)
\begin{equation}
\begin{split}
 \bhat{H}={}&\frac{\Delta}{2}\bhat{\sigma}^{z}
 +\omega_c\sum_n \bhat{a}_n^{\dagger}\bhat{a}_n
 -J\sum_n\left(\bhat{a}_{n+1}^{\dagger}\bhat{a}_n+\hc\right)\\
 &+g\bhat{\sigma}^x\sum_{j=1}^{2}
 \left(e^{\ii\varphi_j}\bhat{a}_{x_j}^{\dagger}
 +e^{-\ii\varphi_j}\bhat{a}_{x_j}\right).
\end{split}
 \label{eq:Hreal}
\end{equation}

The first term is the free Hamiltonian of the emitter, whose ground state $\ket{g}$ and excited state $\ket{e}$ are separated by the bare transition frequency $\Delta$. Here $\bhat{\sigma}^{z}=\ket{e}\!\bra{e}-\ket{g}\!\bra{g}$ and $\bhat{\sigma}^{x}=\bhat{\sigma}^{+}+\bhat{\sigma}^{-}$ are Pauli operators, with $\bhat{\sigma}^{+}=\ket{e}\!\bra{g}=(\bhat{\sigma}^{-})^{\dagger}$ the raising operator.
The second and third terms describe the waveguide. The bosonic operator $\bhat{a}_n$ ($\bhat{a}_n^{\dagger}$) annihilates (creates) a photon in the $n$-th cavity, and $\omega_c$ is the bare cavity frequency. The parameter $J$ is the hopping rate between \chg{nearest-neighbor} cavities, \chg{quantifying} how fast a photon tunnels from one resonator to the next through their mutual coupling, and thus sets both the bandwidth and the propagation speed of excitations along the array.
The last term describes the emitter--waveguide interaction. The emitter dipole couples with strength $g$ to the field of the cavities at generic sites $x_1$ and $x_2$, and $\varphi_1$, $\varphi_2$ are the phases of the two couplings~\cite{Blais,Koch,Wang_2022}. 

Assuming periodic boundary conditions and unit lattice spacing, the waveguide Hamiltonian is invariant under lattice translations and is therefore diagonalized by the plane-wave modes 
\begin{equation}
 \bhat{a}_n=\frac{1}{\sqrt{N}}\sum_k e^{\ii kn}\,\bhat{a}_k,
 \qquad k=\frac{2\pi m}{N}\in(-\pi,\pi],
 \label{eq:Fourier}
\end{equation}
where $\bhat{a}_k$ ($\bhat{a}_k^{\dagger}$) annihilates (creates) a waveguide excitation with momentum $k$. Substituting Eq.~\eqref{eq:Fourier} into Eq.~\eqref{eq:Hreal} and using $\sum_n\bhat{a}_{n+1}^{\dagger}\bhat{a}_n=\sum_k e^{-\ii k}\bhat{a}_k^{\dagger}\bhat{a}_k$, the Hamiltonian becomes
\begin{equation}
 \bhat{H}=\frac{\Delta}{2}\bhat{\sigma}^{z}
 +\sum_k\omega_k \bhat{a}_k^{\dagger}\bhat{a}_k
 +\bhat{\sigma}^{x}\sum_k\left(g_k \bhat{a}_k^{\dagger}+g_k^{*}\bhat{a}_k\right).
 \label{eq:Hk}
\end{equation}
In this representation the waveguide is a collection of independent bosonic modes, with frequencies following the dispersion relation  $ \omega_k=\omega_c-2J\cos k$ .
 
The waveguide thus supports a single band of propagating modes of width $4J$ \chg{centered} at $\omega_c$, with group velocity $v_k=\partial_k\omega_k=2J\sin k$. This velocity vanishes at the band edges and reaches its maximum value $2J$ at the band centre. \chg{Henceforth,} we consider $\omega_c>2J$, so that all photonic frequencies are positive.
The coupling amplitude between the emitter and the mode with momentum $k$ is given by
\begin{equation}
 g_k=\frac{g}{\sqrt{N}}\sum_{j=1}^{2}e^{\ii(\varphi_j-kx_j)} \,\,.
 \label{eq:gkgeneral}
\end{equation}

Only the relative position and the relative phase of the two coupling points are physically relevant. Since the array is invariant under lattice translations, we can shift the origin so that the first coupling point sits at $x_1=0$. Moreover, given that a global phase does not impact the results, we can also impose $\varphi_1=0$.
Therefore, without loss of generality, we define $d=x_2-x_1=x_2$ and $\phi=\varphi_2-\varphi_1=\varphi_2$ as the relative distance and phase between the points, respectively. This yields
\begin{equation}
 |g_k|^2=\frac{2g^2}{N}\left[1+\cos(\phi-kd)\right] \,.
 \label{eq:gksq}
\end{equation}
The coupling strength is modulated by the interference between the two coupling points. Since the phase changes sign with the propagation direction, $|g_k|$ and $|g_{-k}|$ generally differ for $\phi\neq0,\pi$, i.e.\ the emitter couples differently to right- and left-propagating modes. For $\phi=0,\pi$ the couplings are real, time-reversal symmetry is preserved, and $|g_k|=|g_{-k}|$. A nontrivial phase $\phi$ breaks this symmetry and is the origin of chirality~\cite{Ramos2016,DuSuperatoms2025,Almanakly2025,YangMirhosseini2026,PintoValley2026,WangGiantMolecule2026,PhysRevA.109.063709}.

Spin-boson models are fully characterized by their spectral density \cite{ripoll2022quantum,Leggett1987}, which encodes all information about the bath modes and their coupling to the two-level system. To make the directional structure explicit, we decompose the spectral density into contributions from \chg{right- and left-moving} modes, i.e., $\Jtot(\omega)=\JR(\omega)+\JL(\omega)$, where
\begin{equation}
 \mathcal{J}_{R(L)}(\omega)=2\pi\sum_{\substack{v_k>0\\(v_k<0)}}|g_k|^2\,\delta(\omega-\omega_k) \,.
 \label{eq:JRLdef}
\end{equation}
For a frequency $\omega$ within the photonic band, we define $k_\omega = \arccos[(\omega_c-\omega)/2J] \in (0,\pi)$, so that $\pm k_\omega$ correspond to the \chg{right- and left-moving} modes, respectively. Using Eq.~\eqref{eq:gksq} and taking the limit $N\to\infty$, we obtain
\begin{align}
 \mathcal{J}_{R(L)}(\omega)&=\frac{\Jinc(\omega)}{2}\left[1+\cos(\phi\mp k_\omega d)\right],
 \label{eq:JRL}\\
 \Jinc(\omega)&=\frac{4\pi g^2}{N}
 \sum_k\delta(\omega-\omega_k).
 \label{eq:Jinc}
\end{align}
\begin{figure*}[t]
\centering
\includegraphics[width=0.96\textwidth]{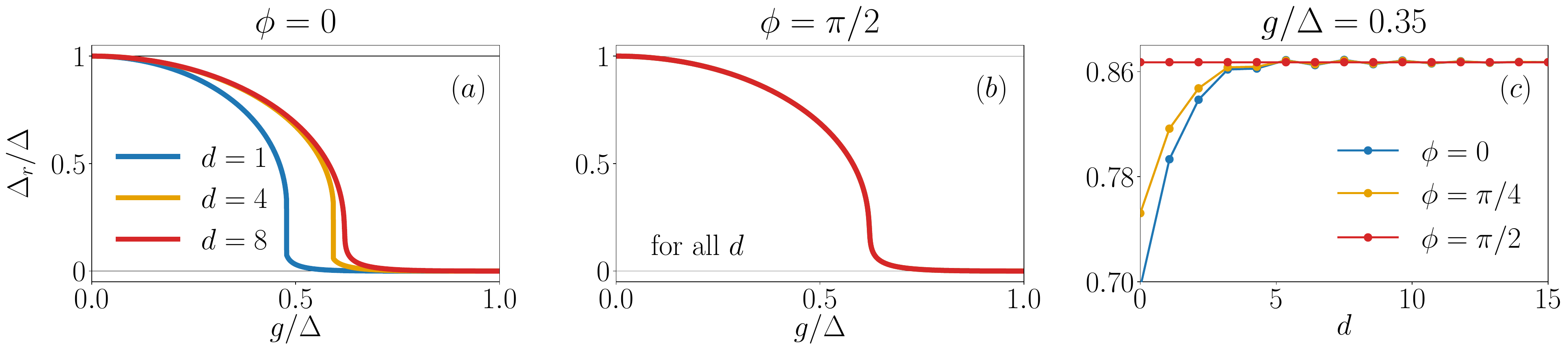}
\caption{Polaron ground-state renormalization for the tight-binding waveguide. (a) $\Delta_r/\Delta$ as a function of the bare coupling for several distances at $\phi=0$. (b) The corresponding curves at $\phi=\pi/2$, where all distances collapse. (c) Distance dependence at $g/\Delta=0.35$ and representative phases. The curves are obtained from Eq.~\eqref{eq:Drpaired} on a grid with $N=1536$, for $\omega_c=1$, $J=0.4$, and $\Delta=0.7$.}
\label{fig:equilibrium}
\end{figure*}
Here $\Jinc$ is the spectral density obtained by neglecting the interference arising from the nonlocal coupling. It is proportional to the density of states of the array and diverges at the band edges \cite{van}. By using the angle addition and subtraction formulae, the sum and the difference of the two directional contributions read
\begin{align}
 \Jtot(\omega)&=\Jinc(\omega)\left[1+\cos\phi\cos(k_\omega d)\right],
 \label{eq:Jsum}\\
 \JR(\omega)-\JL(\omega)&=\Jinc(\omega)\sin\phi\sin(k_\omega d).
 \label{eq:Jdiff}
\end{align}
\chg{Equation}~\eqref{eq:Jsum} is even under $\phi\to-\phi$, whereas Eq.~\eqref{eq:Jdiff} is odd. In the following, we distinguish between \textit{direction-insensitive}
quantities, which depend only on the sum
$\mathcal{J}_{R}(\omega)+\mathcal{J}_{L}(\omega)$, and
\textit{direction-resolved} quantities, which compare the two propagation directions and are therefore sensitive to the difference $\mathcal{J}_{R}(\omega)-\mathcal{J}_{L}(\omega)$. The former include,
for example, the renormalized atomic frequency and the total photon population, whereas the latter include the left-right imbalance of the photon cloud and the directional emission rates.

\section{Polaron theory}
\label{sec:polaron}

To describe the USC regime beyond the rotating-wave approximation, we employ a variational polaron transformation. This unitary transformation dresses the two-level system with a coherent cloud of photons and aims to disentangle the atom from the waveguide. The variational parameters are chosen such that the ground state of the transformed Hamiltonian is as close as possible to $\ket{g}\otimes\ket{0}$. 

The main results of this approach presented in this section are: the form of the variational displacements $f_k$, the renormalized atomic frequency $\Dr$, and the effective number-conserving Hamiltonian $\bhat{H}_P$ governing the low-energy excitations of the system [Eqs.~\eqref{eq:fk}--\eqref{eq:Heff}]. The explicit derivation, namely the transformation of the bosonic and spin operators, the minimization of the variational ground-state energy, and the second-order expansion in the bosonic operators that yields $\bhat{H}_P$, is reported in Appendix~\ref{app:polaron}.

\subsection{Variational ground state}

The variational polaron transformation is defined by the unitary operator~\cite{SilbeyHarris1984,Leggett1987}
\begin{equation}
 \bhat{U}_{P}=\exp\left[-\bhat{\sigma}^x\sum_k
 \left(f_k \bhat{a}_k^{\dagger}-f_k^{*}\bhat{a}_k\right)\right],
 \label{eq:UP}
\end{equation}
where the generally complex amplitudes $f_k$ are determined by minimizing the ground-state energy. The variational ground state is
\begin{equation}
 \ket{\Psi_{\mathrm{GS}}}=\bhat{U}_P\ket{g}\ket{0} \,.
 \label{eq:polaronGSansatz}
\end{equation}
Using the relation $\bhat{U}_P^{\dagger}\bhat{a}_k\bhat{U}_P=\bhat{a}_k-\bhat{\sigma}^x f_k$,
and retaining the vacuum expectation value of the transformed spin operator, one obtains
\begin{equation}
 E_{\mathrm{GS}}=-\frac{\Dr}{2}
 +\sum_k\omega_k|f_k|^2
 -\sum_k\left(g_k f_k^{*}+g_k^{*}f_k\right),
 \label{eq:EGS}
\end{equation}
where
\begin{equation}
 \Dr=\Delta\exp\left[-2\sum_k|f_k|^2\right]
 \label{eq:DeltarDef}
\end{equation}
plays the role of a renormalized transition frequency. Minimization with respect to $f_k^{*}$ gives
\begin{equation}
 f_k=\frac{g_k}{\omega_k+\Dr}.
 \label{eq:fk}
\end{equation}
Equations~\eqref{eq:DeltarDef} and \eqref{eq:fk} lead to the self-consistent polaron equation,
\begin{equation}
 \Dr=\Delta\exp\left[-2\sum_k
 \frac{|g_k|^2}{(\omega_k+\Dr)^2}\right].
 \label{eq:DeltarSC}
\end{equation}
Equivalently, it can be rewritten as
\begin{equation}
 \Dr=\Delta\exp\left[-\frac{1}{\pi}
 \int d\omega\frac{\Jtot(\omega)}{(\omega+\Dr)^2}\right].
 \label{eq:DeltarSpectral}
\end{equation}
The approximation is controlled in the low-energy sector and has been tested against matrix-product-state and numerical diagonalization results in spin-boson and waveguide-QED models \cite{DiazCamacho2016,SanchezBurillo2019,RomanRoche2020,Terradas2022}.

\subsection{Number-conserving low-energy Hamiltonian}

The polaron transformation is what makes a perturbative treatment possible, since, in the original frame, the coupling to the photonic modes in the USC regime cannot be treated perturbatively. In the polaron frame, instead, the dominant part of the interaction is absorbed into the coherent photon cloud dressing the atom\chg{.} The low-energy states then contain only a few excitations above the dressed vacuum $\ket{\Psi_{\mathrm{GS}}}$, so the transformed Hamiltonian can be expanded in powers of the bosonic operators. \chg{At linear order, the variational condition~\eqref{eq:fk} cancels exactly the counter-rotating terms proportional to $\bhat{\sigma}^+\bhat{a}_k^\dagger$ and $\bhat{\sigma}^-\bhat{a}_k$ (see Appendix~\ref{app:polaron}). The remaining terms that do not conserve the excitation number involve at least two bosonic operators and are proportional to $\Dr$; they are therefore suppressed in the low-energy sector,} and a rotating-wave approximation becomes justified in the polaron frame. Expanding $\bhat{H}_P=\bhat{U}_P^{\dagger}\bhat{H}\bhat{U}_P$ to second order in the bosonic operators and discarding the non-number-conserving terms, we obtain (see Appendix~\ref{app:polaron})
\begin{align}
 \bhat{H}_P\simeq{}&E_{\mathrm{GS}}+\Dr\bhat{\sigma}^+\bhat{\sigma}^-
 +\sum_k\omega_k \bhat{a}_k^{\dagger}\bhat{a}_k
 \nonumber\\
 &+\sum_k\left(G_k\bhat{\sigma}^-\bhat{a}_k^{\dagger}
 +G_k^{*}\bhat{\sigma}^+\bhat{a}_k\right)+\Hloc,
 \label{eq:Heff}
\end{align}
where
\begin{equation}
 G_k=2\Dr f_k=\frac{2\Dr g_k}{\omega_k+\Dr}
 \label{eq:Gk}
\end{equation}
are the dressed number-conserving couplings, and
\begin{equation}
 \Hloc=-2\Dr\bhat{\sigma}^z\sum_{k,p}f_k f_p^{*}\bhat{a}_k^{\dagger}\bhat{a}_p
 \label{eq:Vloc}
\end{equation}
is a photonic potential localized on the polaron cloud. Since $\bhat{H}_P$ conserves the number of excitations, its low-energy dynamics can be studied within the single-excitation sector. Therefore, defining \chg{$\ket{e}\equiv\ket{e}\otimes\ket{0}$, $\ket{k}\equiv\ket{g}\otimes\bhat{a}_k^\dagger\ket{0}$, and} the projector $\bhat{\Pi}_1=\dyad{e}+\sum_k\dyad{k}$, the low-energy effective Hamiltonian reads
\begin{align}
\bhat{H}_\mathrm{eff} = {}& \bhat{\Pi}_1\bhat{H}_P\bhat{\Pi}_1 \nonumber \\
 ={}&E_{\mathrm{GS}}\!+\Dr\dyad{e}\!
 +\!\sum_k\omega_k\dyad{k}\! +2\Dr\!\!\sum_{k,p}f_kf_p^{*}\dyad{k}{p}
 \nonumber\\
 &+2\Dr\sum_k\left(f_k\dyad{k}{e}
 +f_k^{*}\dyad{e}{k}\right).
 \label{eq:H1full}
\end{align}
The dressed couplings~\eqref{eq:Gk} define an effective spectral density
\begin{equation}
 \bar{\mathcal J}(\omega)=2\pi\sum_k|G_k|^2\delta(\omega-\omega_k)
 =\frac{4\Dr^2}{(\omega+\Dr)^2}\Jtot(\omega).
 \label{eq:Jpolaron}
\end{equation}
At resonance the dressing factor equals one, so that $\bar{\mathcal J}(\Dr)=\Jtot(\Dr)$. Hence, to lowest order, the spontaneous-emission rate in the polaron frame is given by the bare spectral density evaluated at the renormalized frequency, and USC effects enter mainly through $\Dr$, consistently with previous polaron treatments of spontaneous emission~\cite{DiazCamacho2016,RomanRoche2020,GonzalezGutierrez2021,Terradas2022,tufarelli}.

\section{Equilibrium properties}
\label{sec:equilibrium}

We now use the polaron description to investigate the equilibrium properties of the chiral giant atom. Because the atom couples to the waveguide at two points, each photonic mode $k$ couples to it through the superposition of two contributions, whose interference depends on the coupling phase $\phi$ and on the phase $kd$ accumulated over the separation $d$ between the coupling points. This interference enters all the equilibrium quantities derived above, such as the renormalized frequency $\Dr$, the ground-state energy $E_{\mathrm{GS}}$, and the number of photons in the polaron cloud. In the following, we analyze how these quantities depend on $\phi$ and $d$. We find that the distance-dependent part of the interference is weighted by $\cos\phi$, so that for $\cos\phi=0$ the equilibrium properties become independent of the separation between the coupling points. The general proof of this result, including the case of unequal coupling strengths, is given in Appendix~\ref{app:proof}.

\subsection{Distance-insensitive renormalized transition frequency}

The renormalized frequency $\Dr$ is determined by the total weight of the photonic modes dressing the atom, which for a giant atom results from the interference between the two coupling points. Substituting Eq.~\eqref{eq:gksq} into Eq.~\eqref{eq:DeltarSC}, we obtain
\begin{equation}
 \Dr=\Delta\exp\left[-\frac{4g^2}{N}\sum_k
 \frac{1+\cos(\phi-kd)}{(\omega_k+\Dr)^2}\right].
 \label{eq:Drbeforepair}
\end{equation}
Using $\cos(\phi-kd)=\cos\phi\cos(kd)+\sin\phi\sin(kd)$, the sum splits into an even term and an odd term under $k\to-k$. Since both the dispersion relation and the set of allowed momenta are invariant under this transformation, the contributions proportional to $\sin(kd)$ cancel pairwise, and the self-consistent equation becomes
\begin{equation}
 \Dr=\Delta\exp\left[-\frac{4g^2}{N}\sum_k
 \frac{1+\cos\phi\cos(kd)}{(\omega_k+\Dr)^2}\right].
 \label{eq:Drpaired}
\end{equation}
The coupling phase therefore enters $\Dr$ only through $\cos\phi$, which multiplies the whole distance-dependent part of the sum. In particular, for $\cos\phi=0$ the right-hand side of Eq.~\eqref{eq:Drpaired} no longer depends on $d$, and neither does its self-consistent solution: the renormalized frequency is independent of the separation between the coupling points. Notably, this property does not rely on a large-distance or weak-coupling limit. 
Within the polaron Ansatz, it follows solely from inversion symmetry, and thus holds for arbitrary separation and coupling strengths.

\begin{figure*}[t]
\centering
\includegraphics[width=0.96\textwidth]{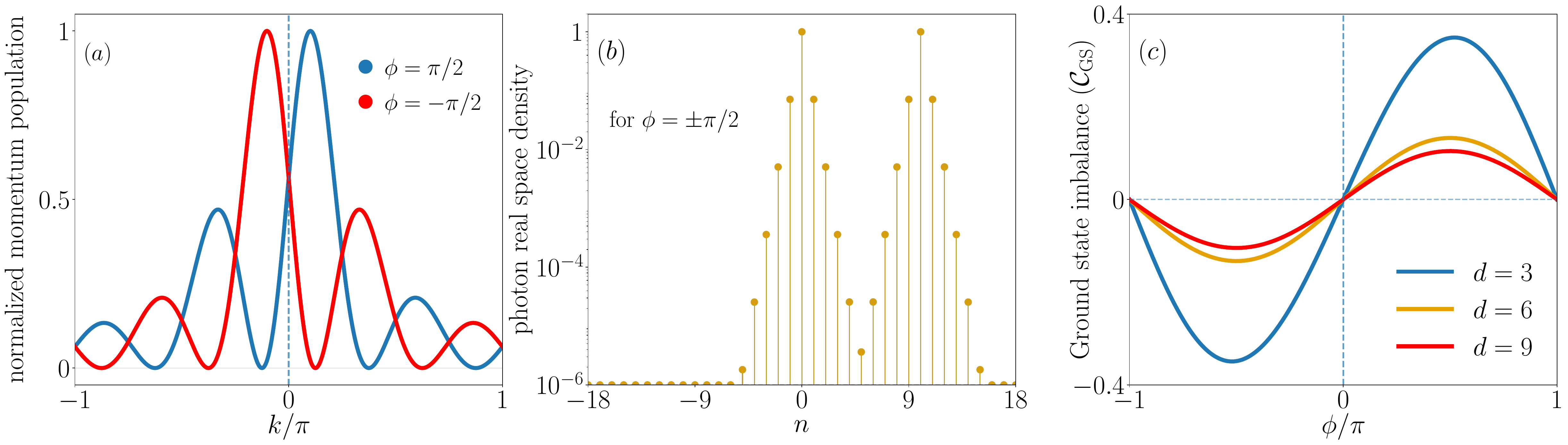}
\caption{Directional content of the virtual-photon cloud.
(a) Momentum occupation for opposite phases at $g/\Delta=0.35$
and \chg{$d=4$}; changing $\phi$ to $-\phi$ interchanges the
populations of $k$ and $-k$.
(b) Static real-space density for $d=10$ and $\phi=\pm \pi/2$.
The local density remains reflection symmetric for equal
coupling magnitudes, although the momentum distribution
is chiral. Other parameters are those of
Fig.~\ref{fig:equilibrium}.
(c) Ground-state imbalance $\mathcal C_{\mathrm{GS}}$
as a function of phase.
}
\label{fig:virtualchirality}
\end{figure*}

\chg{Figure}~\ref{fig:equilibrium} highlights the central role of the relative coupling phase in controlling the interplay between USC and spatially-nonlocal interference. \chg{Figure}~\ref{fig:equilibrium}(a) shows that, for $\phi=0$, the renormalized transition frequency $\Delta_r$ depends appreciably on the separation $d$ between the coupling points. In contrast, Fig.~\ref{fig:equilibrium}(b) demonstrates the collapse of all curves onto a single universal dependence at $\phi=\pi/2$, i.e., the renormalized frequency becomes independent of $d$, even as the system enters the \chg{deep-strong-coupling (DSC)} regime. 
\chg{Figure}~\ref{fig:equilibrium}(c) further clarifies the phase dependence of $\Dr$. For $\phi=0$, in the USC regime, the renormalized frequency exhibits a pronounced distance modulation, whereas intermediate phases progressively suppress this modulation, which vanishes identically at $\phi=\pi/2$.
\chg{This modulation is appreciable only for separations of a few lattice sites. Since $\omega_k+\Dr>0$ across the band, the distance-dependent sum in Eq.~\eqref{eq:Drpaired} decays exponentially with $d$, on the length scale $\kappa^{-1}$ defined by $\cosh\kappa=(\omega_c+\Dr)/2J$; for $g/\Delta=0.35$, $\kappa^{-1}\simeq0.8$ lattice sites.}
\chg{We note that the validity of the polaron approach becomes questionable deep in the USC regime and in the DSC regime, where $\Dr\to0$ may invalidate the RWA in the polaron frame. In the following, we therefore restrict ourselves to coupling strengths $g/\Delta\lesssim0.5$.}

The spectral interpretation of the distance \chg{independence} is particularly transparent. At the special phase,
\begin{equation}
 \Jtot(\omega)\big|_{\phi=\pi/2}=\Jinc(\omega),
 \label{eq:Jspecial}
\end{equation}
for every frequency, not only at resonance. Consequently, any direction-insensitive quantity that depends on the bath through $\Jtot$ alone has the same value for all separations between the coupling points.
This distance independence is not restricted to the polaron
approximation. For each pair of degenerate modes $k$ and $-k$ \chg{($0<k<\pi$)},
we can introduce the bright-mode operator
$\bhat{b}_k^\dagger=
 (g_k\bhat{a}_k^\dagger+g_{-k}\bhat{a}_{-k}^\dagger)
 /(\sqrt{|g_k|^2+|g_{-k}|^2}).$
The orthogonal mode is decoupled from the emitter. Since
$\omega_k=\omega_{-k}$, this rotation leaves the free
waveguide Hamiltonian invariant. At $\phi=\pi/2$,
$|g_k|^2+|g_{-k}|^2=4g^2/N$, independently of $d$.
Thus, the resulting bright-mode Hamiltonian is independent
of the coupling-point separation even before the polaron
transformation. Consequently, the equilibrium properties
and reduced dynamics of the emitter are independent of $d$
for bath states invariant under these mode rotations,
including the vacuum. By contrast, the spatially resolved
photonic observables can retain a dependence on $d$.
The polaron approximation gives explicit expressions for
the ground-state dressing and its directional structure.

\subsection{Distance-insensitive ground-state observables}

The laboratory-frame ground state associated with Eq.~\eqref{eq:polaronGSansatz} can be written as
\begin{equation}
 \ket{\Psi_{\mathrm{GS}}}=\frac{1}{\sqrt{2}}
 \left(\ket{+}\bigotimes_k\ket{-f_k}
 -\ket{-}\bigotimes_k\ket{f_k}\right),
 \label{eq:catGS}
\end{equation}
where $\bhat{\sigma}^x\ket{\pm}=\pm\ket{\pm}$ and $\ket{\pm f_k}$ denote coherent states of mode $k$ with amplitude $\pm f_k$. Equation~\eqref{eq:catGS} is a light--matter Schr\"odinger-cat state, in which the two eigenstates of $\bhat{\sigma}^x$ are entangled with opposite displacements of the photonic field. It makes explicit that, in the USC regime, the ground state contains both a finite atomic excitation and a cloud of virtual photons dressing the emitter. The atomic dressing is quantified by the probability of finding the atom in its excited state,
\begin{equation}
 P_{e,\mathrm{GS}}=\frac{1}{2}
 \left(1-\frac{\Dr}{\Delta}\right).
 \label{eq:PeGS}
\end{equation}
The photonic dressing is instead described by the occupation of each mode,
\begin{equation}
 n_{k,\mathrm{GS}}=\mel{\Psi_{\mathrm{GS}}}{\bhat{a}_k^\dagger \bhat{a}_k}{\Psi_{\mathrm{GS}}}
 =|f_k|^2=\frac{|g_k|^2}{(\omega_k+\Dr)^2},
 \label{eq:nkGS}
\end{equation}
so that the total number of virtual photons is
\begin{equation}
 N_{\mathrm{ph}}^{\mathrm{GS}}=\sum_k|f_k|^2
 =-\frac{1}{2}\ln\frac{\Dr}{\Delta},
 \label{eq:NphGS}
\end{equation}
where the last equality follows from Eq.~\eqref{eq:DeltarDef}. Equations~\eqref{eq:PeGS} and \eqref{eq:NphGS} show that both the atomic and the total photonic dressing are completely determined by $\Dr$. Hence, at $\cos\phi=0$, they are independent of the distance between the two coupling points.
The same holds for the variational ground-state energy. Using Eq.~\eqref{eq:fk}, we obtain
\begin{equation}
 E_{\mathrm{GS}}=-\frac{\Dr}{2}
 -\sum_k|g_k|^2\frac{\omega_k+2\Dr}
 {(\omega_k+\Dr)^2}.
 \label{eq:EGSoptimal}
\end{equation}
As for Eq.~\eqref{eq:Drpaired}, pairing the modes $k$ and $-k$ cancels the term proportional to $\sin\phi\sin(kd)$, so that $E_{\mathrm{GS}}$ is also independent of $d$ at $\cos\phi=0$.

While these previous global quantities are insensitive to the separation between the coupling points, the spatial structure of the photon cloud is not. The real-space displacement of the cloud is
\begin{equation}
 f_n=\frac{1}{\sqrt N}\sum_k f_k e^{\ii kn}.
 \label{eq:fn}
\end{equation}
Introducing the single-coupling-point function
\begin{equation}
 F_m=\frac{1}{N}\sum_k\frac{e^{\ii km}}{\omega_k+\Dr},
 \label{eq:Fm}
\end{equation}
which is real and even in $m$ for an even dispersion relation, we find
\begin{equation}
 f_n=g\left(F_n+e^{\ii\phi}F_{n-d}\right).
 \label{eq:fnF}
\end{equation}
The cloud is therefore the superposition of two clouds centered at the coupling points $n=0$ and $n=d$, with relative phase $\phi$. The corresponding ground-state photon density is
\begin{equation}
 n_{n,\mathrm{GS}}=|f_n|^2
 =g^2\left(F_n^2+F_{n-d}^2
 +2\cos\phi\, F_nF_{n-d}\right).
 \label{eq:realspaceGS}
\end{equation}
The last term describes the interference between the two clouds. At $\cos\phi=0$ it vanishes, and the photon density reduces to the sum of the densities generated by the two coupling points separately. 

\chg{Figure}~\ref{fig:virtualchirality}(a) shows the normalized momentum distribution of virtual photons for $\phi=\pi/2$ and $\phi=-\pi/2$. \chg{As discussed in Sec.~\ref{sec:direction_content} below}, reversing the coupling phase interchanges the populations of counterpropagating modes, thereby reversing the chirality of the photon cloud while leaving its total population unchanged. This behavior reflects the distinct symmetry properties of the total and directional spectral densities under the transformation $\phi\rightarrow-\phi$: the former is even, whereas the latter changes sign. 
\chg{Figure}~\ref{fig:virtualchirality}(b) demonstrates that this momentum-space asymmetry does not entail a corresponding spatial asymmetry. The real-space photon density remains reflection symmetric for both phases, and the two profiles coincide at $\phi=\pm\pi/2$, where the interference contribution to the local density vanishes. Thus, the chirality of the ground state is encoded not in an asymmetric spatial intensity profile, but in the momentum distribution and phase correlations of the counterpropagating modes\chg{.}

\section{Chiral Virtual-Photon Dressing and
			Directional Radiation}
\label{sec:chirality}

In Sec.~\ref{sec:equilibrium}, we showed that, when $\cos\phi=0$, the total ground-state dressing becomes independent of the distance between the coupling points, while some properties of the virtual-photon cloud can still depend on it. 
Here, we characterize the directional structure of \chg{the} momentum distribution of the photon cloud, which can still remain asymmetric between modes propagating in opposite directions, and then show how it can be converted into propagating radiation by suddenly switching off the light--matter coupling.

\subsection{Directional content of the virtual-photon cloud}
\label{sec:direction_content}

We define the right- and
left-moving components of the virtual-photon population
in the dressed ground state as
\begin{equation}
 \chg{N^{\mathrm{GS}}_{R(L)}}=\sum_{\substack{v_k>0 \\ (v_k<0)}}|f_k|^2,
 \label{eq:NRNL}
\end{equation}
and introduce the normalized directional imbalance
$\mathcal C_{\mathrm{GS}}=\chg{(N^{\mathrm{GS}}_R-N^{\mathrm{GS}}_L)/(N^{\mathrm{GS}}_R+N^{\mathrm{GS}}_L)}$.
Using Eq.~\eqref{eq:fk}, we obtain
\begin{equation}
 \mathcal C_{\mathrm{GS}}
 =\!
 \sin\phi\,\!
 \left[
 \sum_{k>0}
 \dfrac{\sin(kd)}{(\omega_k+\Dr)^2}
 \right]\!\!\!
 \left[
 \sum_{k>0}
 \dfrac{1+\cos\phi\cos(kd)}{(\omega_k+\Dr)^2}\right]^{-1}.
 \label{eq:Cgeneral}
\end{equation}
Although the explicit factor $g^2$ cancels between numerator and denominator, the coupling strength still enters implicitly through the renormalized frequency $\Dr$.
An equivalent expression in terms of the directional spectral densities is given by
\begin{equation}
 \mathcal C_{\mathrm{GS}}
 \!
 =
 \!
 \left[
 \int\! \!d\omega
 \dfrac{\JR(\omega)-\JL(\omega)}{(\omega+\Dr)^2}
 \right]
 \!\!\!
 \left[
 \int \!\!d\omega
 \dfrac{\JR(\omega)+\JL(\omega)}{(\omega+\Dr)^2}
 \right]
 ^{-1}\!\!\!.
 \label{eq:Ccontinuum}
\end{equation}
Unlike the directionality of spontaneous emission,
which is determined by the coupling to resonant modes,
Eq.~\eqref{eq:Ccontinuum} involves the entire
photonic band. Therefore, the directional imbalance is a broadband property of the virtual-photon cloud.
\begin{figure*}[t]
\centering
\includegraphics[width=0.96\textwidth]{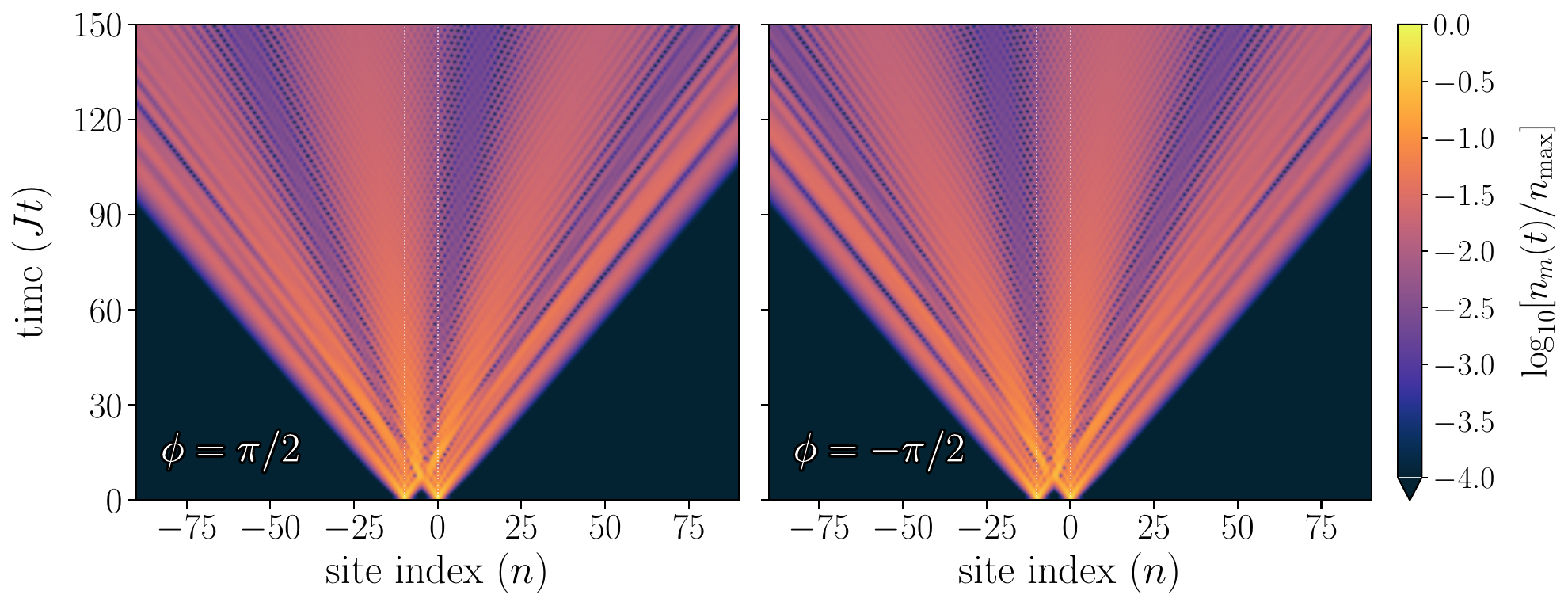}
\caption{Release of the virtual-photon cloud after
a sudden switch-off. The spacetime density \chg{$n_m(t)$}
is calculated from Eq.~\eqref{eq:fnt} for $g/\Delta=0.35$,
$d=10$, and opposite phases. \chg{Time is measured in units of $J^{-1}$, and the dotted vertical lines mark the coupling points (here at $n=-d$ and $n=0$).} The total photon number
is identical for $\phi=\pm\pi/2$, while the
directional imbalance changes sign. Other parameters
are those of Fig.~\ref{fig:equilibrium}.}
\label{fig:quench}
\end{figure*}
At $\phi=\pi/2$, the total virtual-photon population
is independent of $d$, whereas the imbalance
$\mathcal C_{\mathrm{GS}}$ generally is not.
Consequently, the  nonlocal couplings redistribute
a fixed total population between the two propagation
directions. Reversing the phase from $\phi=\pi/2$
to $\phi=-\pi/2$ changes the sign of this imbalance
without modifying $\Dr$, $P_{e,\mathrm{GS}}$,
$N_{\mathrm{ph}}^{\mathrm{GS}}$, or $E_{\mathrm{GS}}$, which are direction-insensitive quantities.
\chg{Figure}~\ref{fig:virtualchirality}(c) shows how \chg{$\mathcal C_{\mathrm{GS}}$} changes sign under $\phi\rightarrow-\phi$ and generally depends on the separation $d$, including at $\phi=\pm\pi/2$. It vanishes when the contributions from the two propagation directions compensate, occurring for $\phi=0$ and $\phi=\pi$ in the inversion-symmetric waveguide considered here. This reveals a fundamental distinction between global and directional properties of the dressed ground state: at $\phi=\pm\pi/2$, the total number of virtual photons is independent of $d$, whereas their partition between right- and left-moving modes remains sensitive to the coupling geometry. The coupling phase therefore provides a means of controlling the chirality of the virtual-photon cloud without modifying its total population.

Importantly, the directional asymmetry does not
contradict the reflection symmetry of the real-space
photon density derived in Eq.~\eqref{eq:realspaceGS}.
For equal coupling strengths, the interference term
in the local density is proportional to $\cos\phi$,
while the difference between the momentum occupations
of opposite propagation directions is proportional
to $\sin\phi$. The ground state can therefore
possess a symmetric spatial photon density and
a chiral momentum distribution simultaneously. 

\begin{figure*}[t]
\centering
\includegraphics[width=0.96\textwidth]{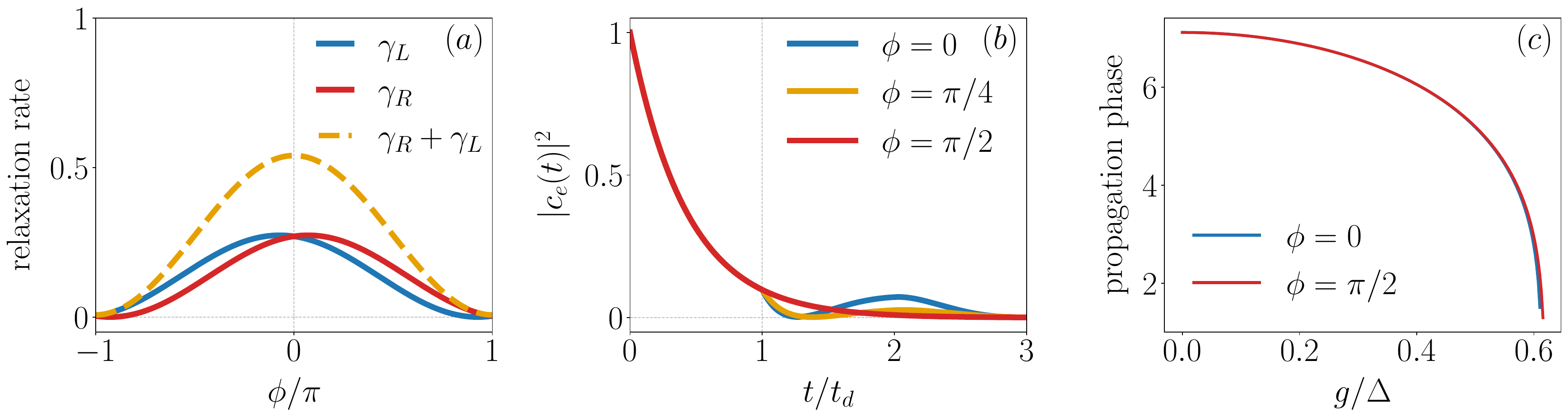}
\caption{Phase-controlled dynamics in the polaron
frame. (a) Right-moving, left-moving, and total
Wigner--Weisskopf rates for $g=0.22$ \chg{($g/\Delta\simeq0.31$)} and $d=6$.
(b) Polaron-frame excited-state survival probability $|c_e(t)|^2$ obtained from
Eq.~\eqref{eq:DDE} for representative phases;
the vertical line marks the first feedback time.
\chg{(c) Propagation phase $k_rd$ as a function of $g/\Delta$ for $\phi=0$ and $\phi=\pi/2$; the renormalization of $\Dr$ shifts the resonant momentum $k_r$ and hence the interference conditions.}
Other parameters are those of Fig.~\ref{fig:equilibrium}.
}
\label{fig:rates}
\label{fig:delay}
\end{figure*}
\subsection{Release of the virtual-photon cloud after a sudden quench}
\label{sec:quench}

The asymmetry discussed above characterizes the
virtual-photon dressing of the interacting ground state.
These photons are not freely propagating excitations.
We now show that a sudden switch-off of the
light--matter coupling converts their momentum
distribution into directional propagating radiation.

Consider the system initially prepared in the dressed
ground state of Eq.~\eqref{eq:catGS}. At $t=0$,
the light--matter coupling is switched off instantaneously.
The state immediately after the quench coincides with
the dressed ground state,
$\ket{\Psi(0^+)}=\ket{\Psi_{\mathrm{GS}}}$.
Tracing out the emitter leaves an equal mixture
of two multimode coherent states with amplitudes
$\pm f_k$. Hence,
\begin{equation}
 \mel{\Psi_{\mathrm{GS}}}
 {\bhat{a}_k^\dagger\bhat{a}_k}
 {\Psi_{\mathrm{GS}}}
 =|f_k|^2.
 \label{eq:quenchNk}
\end{equation}

Under the free waveguide Hamiltonian \chg{$\bhat{H}_{\mathrm w}=\sum_k\omega_k\bhat{a}_k^\dagger\bhat{a}_k$},
the coherent amplitudes evolve as
\begin{equation}
 f_n(t)=\frac{1}{\sqrt N}\sum_k
 f_k e^{\ii kn-\ii\omega_kt},
 \label{eq:fnt}
\end{equation}
and the real-space photon density is $n_m(t)=|f_m(t)|^2$.
The total photon number is conserved after the
quench and equals $N_{\mathrm{ph}}^{\mathrm{GS}}$. The energy carried away by these photons is supplied by the external switching operation; the interacting ground state does not radiate in the absence of the quench.
In an infinite waveguide, or before finite-size
recurrences in a sufficiently large array, the
components with positive and negative group velocities
separate spatially. Their integrated photon numbers
are precisely \chg{$N_R^{\mathrm{GS}}$ and $N_L^{\mathrm{GS}}$} in Eq.~\eqref{eq:NRNL}.
The directionality of the released radiation is
therefore determined by $\mathcal C_{\mathrm{GS}}$.

At $\phi=\pi/2$, the total number of photons released
by the quench is independent of $d$, while their
distribution between the two propagation directions
remains distance dependent.
\chg{Figure}~\ref{fig:quench} shows that reversing the coupling phase from $\phi=\pi/2$ to $\phi=-\pi/2$ reverses the directionality of the propagating photon cloud. The space–time patterns exhibit opposite directional imbalances, while the total photon population remains unchanged. This provides a direct dynamical signature of the ground-state chirality discussed in Fig.~\ref{fig:virtualchirality}: although the two configurations contain the same total number of virtual photons, their release produces radiation propagating preferentially in opposite directions.
Since the photons are released by switching off the interaction, rather than through spontaneous relaxation, the protocol allows directional radiation to be generated on the timescale of the quench, and their subsequent propagation is governed by the waveguide group velocity.

\section{Spontaneous emission and delayed feedback}
\label{sec:decay}
\label{sec:ww}

We now study the dynamics of an initially excited giant atom. Unlike the ground-state switch-off
protocol discussed in Sec.~\ref{sec:chirality},
spontaneous emission is mainly governed by the coupling
between the excited emitter and the photonic continuum.
The effective Hamiltonian $\bhat{H}_\mathrm{eff}$ derived in Sec.~\ref{sec:polaron}
allows us to analytically investigate how USC
modifies both the directional emission rates and
the delayed self-interaction of the chiral giant atom.

Throughout this section, we neglect the local
photonic potential $\Hloc$ in Eq.~\eqref{eq:Heff}.
The resulting one-excitation Hamiltonian has the
form of a Fano--Anderson model, with renormalized
atomic frequency $\Dr$ and dressed couplings $G_k$ \cite{fano,Anderson,PhysRevA.89.053826}. 
Retaining $\Hloc$ introduces an additional rank-one
	scattering potential in the photonic sector, which
	modifies the spectral self-energy and can shift the
	bound-state energies and linewidths. Neglecting this
	term yields the simpler Fano--Anderson Hamiltonian
	used in the following dynamical calculations.
	The corresponding one-excitation spectrum, including
	the effect of $\Hloc$, is discussed in
	Appendix~\ref{app:Vloc}. 

\subsection{Total and directional relaxation rates}
Assuming that $\Dr$ lies inside the photonic band,
the partial spontaneous-emission rates into the
right- and left-propagating waveguide modes are
obtained from Fermi's golden rule,
\begin{equation}
 \gamma_{R(L)}=
 2\pi\sum_{\substack{v_k>0\\(v_k<0)}}
 |G_k|^2\delta(\omega_k-\Dr)
 =\mathcal{J}_{R(L)}(\Dr).
 \label{eq:directionalrates}
\end{equation}
The second equality follows from Eq.~\eqref{eq:Gk},
since the dressed coupling satisfies $G_k=g_k$
at resonance. Let $k_r>0$ satisfy $\omega_{k_r}=\Dr$,
and define the reference rate
$\Gamma_r=\Jinc(\Dr)$.
Using \chg{Eq.~\eqref{eq:JRL}},
we obtain
\begin{align}
 \gamma_{R(L)}&=\frac{\Gamma_r}{2}
 [1+\cos(\phi \mp k_rd)].
\end{align}
The different interference phases reflect the
opposite propagation directions of the emitted
photons. Their sum gives the total relaxation rate
\begin{equation}
 \gamma_r=\gamma_R+\gamma_L
 =\Gamma_r[1+\cos\phi\cos(k_rd)].
\end{equation}
The interference conditions are thus evaluated
at the renormalized frequency $\Dr$, rather
than at the bare transition frequency $\Delta$.
In the USC regime, this produces
a coupling-dependent shift of the interference
conditions.

The normalized directional imbalance is
\begin{equation}
 \mathcal D_r=
 \frac{\gamma_R-\gamma_L}{\gamma_R+\gamma_L}
 =\frac{\sin\phi\sin(k_rd)}
 {1+\cos\phi\cos(k_rd)},
\end{equation}
whenever the total rate is nonzero.
At the special phase $\phi=\pi/2$,
\begin{equation}
 \gamma_r=\Gamma_r,
 \qquad
 \mathcal D_r=\sin(k_rd).
 \label{eq:decayspecial}
\end{equation}
The total relaxation rate is therefore
independent of the distance between the
coupling points, whereas the directionality
remains distance dependent. In particular,
perfectly unidirectional emission is obtained
for $k_rd=(2m+1)\pi/2$, with the propagation
direction determined by the \chg{parity of} $m$. 
These rates characterize the resonant coupling to the two propagation channels. When the
propagation delay is appreciable, the complete
time evolution must instead be obtained from the non-Markovian equations derived below. This result parallels the behavior of the
ground-state virtual-photon population discussed
in Sec.~\ref{sec:chirality}: the total quantity
becomes independent of the distance between coupling points,
while its directional components remain tunable.
However, the two directionalities probe different
properties of the system. The ground-state
imbalance $\mathcal C_{\mathrm{GS}}$ samples the
entire photonic band, whereas $\mathcal D_r$
is determined by the dressed coupling at
the resonant frequency $\Dr$.

Figure~\ref{fig:rates}(a) shows the directional rates as functions of
$\phi$ for $g=0.22$ \chg{($g/\Delta\simeq0.31$)} and $d=6$. The two rates coincide at $\phi=0$ and
$\phi=\pm\pi$, where $\mathcal D_r=0$ and the emission is bidirectional.
For these parameters $\cos(k_rd)$ is close to unity, so the total rate
approaches $2\Gamma_r$ at $\phi=0$, where the contributions from the two
coupling points interfere constructively, and is strongly suppressed
near $\phi=\pm\pi$, where they interfere destructively. At
$\phi=\pm\pi/2$ the total rate reduces to $\Gamma_r$, in agreement with
Eq.~\eqref{eq:decayspecial}, while the residual imbalance between
$\gamma_R$ and $\gamma_L$ is set by $\sin(k_rd)$, which is small for the
chosen distance. Therefore, the emission is only partially directional at
these phases, while perfect unidirectionality would require tuning $d$ such
that $k_rd=(2m+1)\pi/2$.

\subsection{Non-Markovian dynamics and memory kernel}

The rates obtained above describe the resonant
coupling to the photonic continuum. The importance of spectral poles and reservoir memory for giant-atom dynamics has also been discussed in Ref.~\cite{WuAn2026}. To investigate
the dynamics beyond this approximation, we now
derive the memory equation for the excited-state amplitude within the effective polaron Hamiltonian after neglecting $\Hloc$.

Considering the interaction picture with respect to $\Dr\dyad{e}
 +\sum_k\omega_k\dyad{k}$,
we can write the generic single-excitation state as $\ket{\psi(t)}_P=c_e(t)\ket{e}
 +\sum_k c_k(t)\ket{k}$.
Imposing the Schr\"odinger equation, the equations of motion of the atomic and photonic coefficients are
\begin{align}
 \dot c_e(t)&=-\ii\sum_kG_k^*
 e^{-\ii(\omega_k-\Dr)t}c_k(t),
 \label{eq:ceeq}\\
 \dot c_k(t)&=-\ii G_k
 e^{\ii(\omega_k-\Dr)t}c_e(t).
 \label{eq:ckeq}
\end{align}
We choose $c_e(0)=1$ and $c_k(0)=0$, corresponding to an initially excited bare state in the polaron frame (and therefore a dressed initial state in the laboratory frame). The quantity $|c_e(t)|^2$ is the survival probability in the polaron frame, not, in general, the laboratory-frame atomic excitation probability; see Appendix~\ref{app:observables}.
Integrating Eq.~\eqref{eq:ckeq} and substituting
the result into Eq.~\eqref{eq:ceeq}, we obtain
\begin{equation}
 \dot c_e(t)=-\int_0^t d\tau\,
 K(\tau)c_e(t-\tau),
 \label{eq:memoryeq}
\end{equation}
where the memory kernel is
\begin{align}
 K(\tau)
 &=\frac{2\Dr^2}{\pi}\int d\omega
 \frac{\Jtot(\omega)}{(\omega+\Dr)^2}
 e^{-\ii(\omega-\Dr)\tau}.
 \label{eq:Kspectral}
\end{align}
Equation~\eqref{eq:memoryeq} retains the
frequency dependence of the photonic continuum
and the interference associated with the
spatially separated coupling points.

Using Eq.~\eqref{eq:Jsum}, the memory kernel
can be separated into $K(\tau)=K_0(\tau)+\cos\phi\,K_d(\tau)$,
where
\begin{align}
 K_0(\tau)&=\frac{2\Dr^2}{\pi}
 \int d\omega\,
 \frac{\Jinc(\omega)}{(\omega+\Dr)^2}
 e^{-\ii(\omega-\Dr)\tau},
 \label{eq:K0}\\
 K_d(\tau)&=\frac{2\Dr^2}{\pi}
 \int d\omega\,
 \frac{\Jinc(\omega)\cos(k_\omega d)}
 {(\omega+\Dr)^2}
 e^{-\ii(\omega-\Dr)\tau}.
 \label{eq:Kd}
\end{align}
The first term describes the reservoir contribution
that would remain in the absence of interference
between the coupling points. The second contains
the distance-dependent interference contribution.
The latter is multiplied by $\cos\phi$, and consequently, at $\phi=\pi/2$,
\begin{equation}
 K(\tau)\big|_{\phi=\pi/2}=K_0(\tau).
 \label{eq:Kspecial}
\end{equation}
Since $\Dr$ is itself independent of $d$ at
this phase, the entire memory kernel becomes
independent of the distance between the coupling points.

Importantly, this result does not require
a Markov approximation. The cancellation of
the distance-dependent contribution holds for
the full finite-band kernel within the
effective Hamiltonian considered here.
Nevertheless, $K_0(\tau)$ can still be nonlocal
in time because of the finite bandwidth of
the waveguide and the frequency dependence
of the dressed couplings. The absence of
distance-dependent feedback therefore does
not necessarily imply purely exponential decay.

\subsection{Wigner--Weisskopf approximation and delayed feedback}

We now consider the broad-band, linear-dispersion
approximation, assuming the renormalized transition frequency lies sufficiently far from the band edges, which allows the memory equation
to be reduced to a delay differential equation.
To retain the propagation time between the
coupling points while neglecting the short
reservoir memory, we linearize the dispersion
around the resonant momenta,
\begin{equation}
 k_\omega\simeq\kr+(\omega-\Dr)/\vg,
 \qquad
 \td=d/\vg,
 \label{eq:linearization}
\end{equation}
where $v_g=2J\sin k_r$, and replace the slowly varying factor
$4\Dr^2\Jinc(\omega)/(\omega+\Dr)^2$
by its resonant value.

The frequency integral in $K_d$ then produces
a delta function at the propagation time $\td$,
and Eq.~\eqref{eq:memoryeq} becomes
\begin{equation}
 \dot c_e(t)=-\frac{\Gamma_r}{2}c_e(t)
 -\frac{\Gamma_r}{2}\cos\phi\,
 e^{\ii\kr d}c_e(t-\td)\Theta(t-\td),
 \label{eq:DDE}
\end{equation}
where $\Theta(t)$ is the Heaviside step function.
The derivation is detailed in
Appendix~\ref{app:delay}.

Equation~\eqref{eq:DDE} has the same structure as the
weak-coupling delay equation for a chiral giant atom,
but describes the dynamics in the polaron frame.
Within the present approximation, the effects of
USC enter through the renormalized
frequency $\Dr$, which modifies the resonant momentum
$\kr$, the decay rate $\Gamma_r=\Jinc(\Dr)$, and
the propagation time $\td$.
\chg{Figure~\ref{fig:rates}(c) shows how the propagation phase $\kr d$ is modified by the renormalization of the resonant momentum: since $\Dr$ decreases with $g/\Delta$, so do $\kr$ and $\kr d$. The curves for $\phi=0$ and $\phi=\pi/2$ are almost indistinguishable because, for $d=6$, the distance-dependent contribution to Eq.~\eqref{eq:Drpaired} is exponentially small. The shift therefore originates from the polaron renormalization of the transition frequency, and not from the phase-dependent interference between the coupling points.}

At $\cos\phi=0$, all delayed contributions
vanish and the solution reduces to $c_e(t)=e^{-\Gamma_rt/2}$,
independently of $\td$.
This result is exact within the broad-band delay equation,
even when the propagation time is comparable
to or larger than the spontaneous-emission
timescale.
This behavior is illustrated in Fig.~\ref{fig:delay}(b).
Before the first feedback time, the excited-state
population decays with the rate $\Gamma_r$, which
depends only weakly on $\phi$ for the parameters shown.
At later times, increasing $\phi$ from $0$ to $\pi/2$
progressively suppresses the delayed revivals, whose
amplitude is governed by $\cos\phi$, the term ultimately responsible for the discontinuity at $t=t_d$.
\chg{As a result, at later times the excited-state population at $\phi=\pi/2$ can be smaller than at $\phi=0$, even though the corresponding golden-rule total decay rate is smaller [Fig.~\ref{fig:delay}(a)].} 

The distinction between the two sources of
memory is essential. At $\phi=\pi/2$,
the delayed self-interaction associated with
the spatial separation of the coupling points disappears.
However, finite-band reservoir memory is not
eliminated by this cancellation. In the complete
finite-band description of
Eq.~\eqref{eq:memoryeq}, the excited-state
amplitude can still exhibit nonexponential
corrections, although these remain independent
of $d$ within the same approximation.

\section{Conclusions}
\label{sec:conclusions}

We have investigated a chiral giant atom in the
USC regime using a variational
polaron approach.
The central result is the separation between the total
and directional contributions to the atom--waveguide
coupling.
At $\phi=\pi/2$, the total spectral density
is independent of the distance between the coupling
points, an exact property of the model for an
inversion-symmetric waveguide.
Within the polaron description, this independence
extends to the renormalized transition frequency,
the ground-state energy, and the total matter and
photonic dressing.

The directional properties of the ground state,
however, remain dependent on the distance between
the coupling points.
In particular, the virtual-photon cloud has an
asymmetric momentum distribution that can be
converted into directional propagating radiation
by suddenly switching off the coupling.
The total photon number released by the quench
is independent of the distance between the coupling
points at $\phi=\pi/2$, while its directional
distribution can be controlled by changing that distance.
Analogous results could be obtained by periodically and non-adiabatically modulating the coupling strength, which may facilitate experimental implementation (see, e.g., \cite{Wilson2011}).
We have also examined spontaneous emission in the
polaron frame.
After neglecting the local photonic potential,
the memory kernel becomes distance independent
at the same phase.
In the broad-band limit, this eliminates delayed
self-interaction and gives exponential decay with
renormalized parameters, whereas finite-band
reservoir memory can still produce nonexponential
dynamics.
Thus, the cancellation of propagation-induced
feedback persists in the USC
description, while the photonic dressing retains
a nontrivial directional structure.

\section*{Acknowledgments}
	F.R.~acknowledges financial support by the European Union-Next Generation EU with the project ``Quantum Optics in Many-Body photonic Environments'' (QOMBE) code SOE2024\_0000084-CUP B77G24000480006.
    The authors acknowledge the use of ChatGPT-6, ChatGPT-5.6 Sol (OpenAI) and Claude Code Opus 5.5 for assistance with language editing, proofreading, and stylistic improvements. All AI-assisted content was reviewed and revised by the authors, who take full responsibility for the final manuscript.

\appendix

\section{Complex polaron transformation}
\label{app:polaron}

This Appendix gives the main steps leading to Eqs.~\eqref{eq:fk}--\eqref{eq:Heff}. Let us define
\begin{equation}
 \bhat{S}=-\bhat{\sigma}^x\sum_k(f_k\bhat{a}_k^\dagger-f_k^*\bhat{a}_k),
 \qquad \bhat{U}_P=e^{\bhat{S}}.
 \label{eq:Sdef}
\end{equation}
Since $[\bhat{S},\bhat{a}_k]=\bhat{\sigma}^xf_k$ and higher commutators vanish,
\begin{equation}
 \bhat{U}_P^\dagger \bhat{a}_k\bhat{U}_P=\bhat{a}_k-\bhat{\sigma}^xf_k.
 \label{eq:aTransformApp}
\end{equation}
Furthermore, the spin operator can be transformed by using $\{\bhat{\sigma}^z,\bhat{\sigma}^x\}=0$. With $\bhat{A}=\sum_k(f_k\bhat{a}_k^\dagger-f_k^*\bhat{a}_k)$, one obtains
\begin{equation}
 \bhat{U}_P^\dagger\bhat{\sigma}^z\bhat{U}_P=e^{2\bhat{\sigma}^x\bhat{A}}\bhat{\sigma}^z \, ,
 \label{eq:sigmazexact}
\end{equation}
and normal ordering gives
\begin{equation}
 e^{2\bhat{\sigma}^x\bhat{A}}=e^{-2\sum_k|f_k|^2}
 e^{2\bhat{\sigma}^x\sum_kf_k\bhat{a}_k^\dagger}
 e^{-2\bhat{\sigma}^x\sum_kf_k^*\bhat{a}_k} \,.
 \label{eq:normalorder}
\end{equation}
Therefore, the vacuum expectation value is
\begin{equation}
 \mel{0}{\bhat{U}_P^\dagger\bhat{\sigma}^z\bhat{U}_P}{0}
 =e^{-2\sum_k|f_k|^2}\bhat{\sigma}^z \,,
 \label{eq:sigmazvac}
\end{equation}
which defines $\Dr$ as in Eq.~\eqref{eq:DeltarDef}.

Since the variational ground state in the laboratory
	frame is
	$\ket{\Psi_{\mathrm{GS}}}=\hat U_P\ket{g,0}$,
	its energy can equivalently be evaluated as
	\begin{equation}
		E[f]=
		\mel{g,0}{\hat H_P}{g,0}
		=
		\mel{\Psi_{\mathrm{GS}}}{\hat H}{\Psi_{\mathrm{GS}}}.
	\end{equation}
	This gives
\begin{equation}
 E[f]= -\frac{\Delta}{2}e^{-2\sum_k|f_k|^2}
 +\sum_k\omega_k|f_k|^2
 -\sum_k(g_kf_k^*+g_k^*f_k).
 \label{eq:EfunctionalApp}
\end{equation}
Treating $f_k$ and $f_k^*$ as independent variables,
\begin{equation}
 \frac{\partial E}{\partial f_k^*}
 =(\omega_k+\Dr)f_k-g_k=0,
 \label{eq:variationApp}
\end{equation}
which gives Eq.~\eqref{eq:fk}.

To obtain the low-energy Hamiltonian, we expand Eq.~\eqref{eq:sigmazexact} to second order in bosonic operators after normal ordering\chg{, omitting the terms proportional to $\bhat{a}_k^\dagger\bhat{a}_p^\dagger$ and $\bhat{a}_k\bhat{a}_p$, which do not conserve the excitation number},
\begin{align}
 \bhat{U}_P^\dagger\bhat{\sigma}^z\bhat{U}_P\simeq
 \frac{\Dr}{\Delta}\Bigg[1
 &+2\bhat{\sigma}^x\sum_k(f_k\bhat{a}_k^\dagger-f_k^*\bhat{a}_k)
 \nonumber\\
 &-4\sum_{k,p}f_kf_p^*\bhat{a}_k^\dagger \bhat{a}_p\Bigg]\bhat{\sigma}^z.
 \label{eq:sigmazexpansion}
\end{align}
Combining this expression with Eq.~\eqref{eq:aTransformApp}, using the stationarity condition~\eqref{eq:variationApp}\chg{, which cancels exactly the linear counter-rotating terms $\bhat{\sigma}^+\bhat{a}_k^\dagger$ and $\bhat{\sigma}^-\bhat{a}_k$}, and retaining number-conserving terms yields
\begin{align}
 \bhat{H}_P\simeq{}&E_{\mathrm{GS}}+\Dr\bhat{\sigma}^+\bhat{\sigma}^-
 +\sum_k\omega_k\bhat{a}_k^\dagger \bhat{a}_k
 \nonumber\\
 &+2\Dr\sum_k(f_k\bhat{\sigma}^-\bhat{a}_k^\dagger
 +f_k^*\bhat{\sigma}^+\bhat{a}_k)
 \nonumber\\
 &-2\Dr\bhat{\sigma}^z\sum_{k,p}f_kf_p^*\bhat{a}_k^\dagger \bhat{a}_p,
 \label{eq:HeffApp}
\end{align}
which is Eq.~\eqref{eq:Heff}. Terms with three or more bosonic operators and terms changing the excitation number have been discarded consistently with the low-energy polaron approximation.

\section{Distance independence at \texorpdfstring{$\phi=\pi/2$}{phi = pi/2}}
\label{app:proof}

In this Appendix we show that, for $\phi=\pi/2$, all sums over the photonic modes of the form $\sum_k|g_k|^2W_k$, with $W_k$ being a generic even function of $k$, are independent of the separation $d$ between the coupling points. This includes, in particular, any function of the mode frequency, $W_k=F(\omega_k)$, since $\omega_k=\omega_{-k}$. We start from the case of equal coupling strengths, for which
\begin{equation}
 |g_k|^2=\frac{2g^2}{N}\left[1+\cos(\phi-kd)\right].
 \label{eq:gkequal}
\end{equation}
Using $\cos(\phi-kd)=\cos\phi\cos(kd)+\sin\phi\sin(kd)$, we obtain
\begin{equation}
 \sum_k\!|g_k|^2W_k
 \!=\!\frac{2g^2}{N}\!\sum_k\!\left[1\!+\!\cos\phi\cos(kd)\!+\!\sin\phi\sin(kd)\right]W_k.
 \label{eq:expansion}
\end{equation}
Since the set of allowed wavenumbers is symmetric under $k\to-k$, the modes can be grouped in pairs $(k,-k)$. Within each pair, $\cos(kd)W_k$ is even while $\sin(kd)W_k$ is odd, so the contributions of the last term cancel pairwise, leading to $\sum_k\sin(kd)W_k=0$. Therefore
\begin{equation}
 \sum_k|g_k|^2W_k
 =\frac{2g^2}{N}\sum_k\left[1+\cos\phi\cos(kd)\right]W_k.
 \label{eq:evenidentity}
\end{equation}
The only dependence on $d$ is contained in the term proportional to $\cos\phi$. At $\phi=\pi/2$ this term vanishes and
\begin{equation}
 \sum_k|g_k|^2W_k
 =\frac{2g^2}{N}\sum_kW_k,
 \label{eq:evenidentityspecial}
\end{equation}
which is independent of $d$.
Choosing $W_k=(\omega_k+\Dr)^{-2}$ shows directly that the sum entering Eq.~\eqref{eq:Drpaired} is independent of $d$ at $\cos\phi=0$. Other choices reproduce the corresponding statements for the ground-state energy, total photon number, memory kernel, and total decay rate. The result fails if inversion symmetry is broken or if the mode set does not contain the pairs $k$ and $-k$. Equal coupling-point strengths are not required for the cancellation of the distance-dependent term.

For unequal local couplings $g_1$ and $g_2$,
\begin{equation}
 |g_k|^2=\frac{1}{N}
 [g_1^2+g_2^2+2g_1g_2\cos(\phi-kd)].
 \label{eq:unequalg}
\end{equation}
The distance-dependent term is still proportional to $\cos\phi$ after pairing $\pm k$. Therefore the cancellation at $\cos\phi=0$ remains valid, although the baseline spectral density becomes proportional to $g_1^2+g_2^2$ and perfect resonant directionality generally requires $g_1=g_2$.

\section{Derivation of the delay equation}
\label{app:delay}

Starting from Eq.~\eqref{eq:Kspectral}, we write the interference part of the spectral density as
\begin{equation}
 \Jtot(\omega)=\Jinc(\omega)
 \left[1+\frac{\cos\phi}{2}
 (e^{\ii k_\omega d}+e^{-\ii k_\omega d})\right].
 \label{eq:Jexponential}
\end{equation}
Near resonance,
\begin{equation}
 e^{\pm\ii k_\omega d}
 \simeq e^{\pm\ii\kr d}
 e^{\pm\ii(\omega-\Dr)\td}.
 \label{eq:phaseLinear}
\end{equation}
We extend the detuning integral to the real axis and replace the smooth dressed prefactor by its value at resonance:
\begin{equation}
 \frac{2\Dr^2}{\pi}
 \frac{\Jinc(\omega)}{(\omega+\Dr)^2}
 \simeq\frac{\Gamma_r}{2\pi}.
 \label{eq:flatprefactor}
\end{equation}
Using the \chg{identity} $\int_{-\infty}^{\infty}d\nu e^{-\ii\nu(\tau-\td)}= 2 \pi \delta(\tau-\td)$, we obtain, for $\tau\geq0$,
\begin{equation}
 K(\tau)\simeq\frac{\Gamma_r}{2}\delta_+(\tau)
 +\frac{\Gamma_r}{2}\cos\phi\,
 e^{\ii\kr d}\delta(\tau-\td).
 \label{eq:Kdelta}
\end{equation}
Here $\delta_+$ is the one-sided delta distribution, normalized so that its convolution on $[0,t]$ gives the local term without any additional factor. Substitution into Eq.~\eqref{eq:memoryeq} gives Eq.~\eqref{eq:DDE}. The term proportional to $e^{-\ii k_\omega d}$ produces $\delta(\tau+\td)$, which has no support in the causal integration interval $\tau\geq0$ for $\td>0$, and thus results in a zero contribution.

For a tight-binding dispersion, Eq.~\eqref{eq:linearization} is valid when the emitted spectrum is narrow and $v_g\neq0$. Close to a band edge the group velocity vanishes, the linearization fails, and the complete memory kernel must be retained.

\section{Bound states and the local photonic potential}
\label{app:Vloc}
\label{sec:bound}

In this Appendix, we examine the one-excitation
bound states of the effective polaron Hamiltonian.
We first neglect the local photonic potential
$\Hloc$ and derive the corresponding bound-state
and bound-state-in-the-continuum conditions.
We then include $\Hloc$ and show how its
rank-one structure modifies the spectral
pole equation.
All one-excitation energies below are measured relative to the additive constant $E_{\mathrm{GS}}$ in Eq.~\eqref{eq:H1full}.

\subsection{Bound states without the local potential}

Neglecting $\Hloc$ in Eq.~\eqref{eq:H1full},
we consider one-excitation eigenstates of the form
\begin{equation}
 \ket{\Psi_E}=u_e\ket{e}+\sum_k u_k\ket{k}.
 \label{eq:boundansatz}
\end{equation}
The stationary Schr\"odinger equation gives
\begin{align}
 (E-\Dr)u_e&=\sum_kG_k^*u_k,
 \label{eq:boundeq1}\\
 (E-\omega_k)u_k&=G_ku_e.
 \label{eq:boundeq2}
\end{align}
Eliminating the photonic amplitudes, $u_k=G_ku_e/(E-\omega_k)$, leads to the pole equation
\begin{equation}
 E-\Dr-\Sigma(E)=0,
 \label{eq:poleeq}
\end{equation}
where
\begin{equation}
 \Sigma(z)=\sum_k\frac{|G_k|^2}{z-\omega_k}
 =\frac{1}{2\pi}\int d\omega
 \frac{\bar{\mathcal J}(\omega)}{z-\omega}.
 \label{eq:selfenergy}
\end{equation}

For a real solution outside the photonic band,
Eq.~\eqref{eq:poleeq} describes an ordinary
band-edge bound state. Its atomic weight is
\begin{equation}
 |u_e|^2=
 \left[1+\sum_k\frac{|G_k|^2}
 {(E-\omega_k)^2}\right]^{-1}.
 \label{eq:boundnorm}
\end{equation}
The coupling phase modifies both the pole
position and the localization length through
the spectral density $\Jtot$. USC enters through the renormalized
frequency $\Dr$ and the dressed coupling
factor $2\Dr/(\omega+\Dr)$.

For an energy $E$ inside the photonic band,
the retarded self-energy takes the form
\begin{equation}
 \Sigma(E+\ii0^+)=
 \chg{\Lambda(E)}-\frac{\ii}{2}\bar{\mathcal J}(E),
 \label{eq:SigmaImag}
\end{equation}
where
\begin{equation}
 \chg{\Lambda(E)}=\frac{1}{2\pi}
 \mathcal P\int d\omega\,
 \frac{\bar{\mathcal J}(\omega)}{E-\omega}
 \label{eq:boundLambshift}
\end{equation}
is the real part of the retarded self-energy.
A bound state in the continuum (BIC) requires
both a vanishing decay channel and the
fulfillment of the real pole equation:
\begin{align}
 \bar{\mathcal J}(E)&=0,
 \label{eq:BICzero}\\
 E-\Dr-\chg{\Lambda(E)}&=0.
 \label{eq:BICreal}
\end{align}

Since the prefactor in Eq.~\eqref{eq:Jpolaron}
is nonzero for $E>0$, the first condition
is equivalent to
\begin{equation}
 1+\cos\phi\cos(k_Ed)=0,
 \qquad
 \omega_{k_E}=E.
 \label{eq:BICphase}
\end{equation}
For two coupling points of equal strength,
Eq.~\eqref{eq:BICphase} can be satisfied
only when $|\cos\phi|=1$. Explicitly,
\begin{align}
 \phi&=2m\pi,
 &k_Ed&=(2n+1)\pi,
 \label{eq:BIC0}\\
 \phi&=(2m+1)\pi,
 &k_Ed&=2n\pi,
 \label{eq:BICpi}
\end{align}
with integers $m,n$.

These conditions determine the interference
zeros of the coupling to the continuum.
The additional real-energy condition
in Eq.~\eqref{eq:BICreal} must also be
satisfied for a genuine BIC to exist.
The renormalization of $\Dr$ and the
frequency dependence of the dressed
couplings modify this condition with
respect to the weak-coupling regime.

At $\phi=\pi/2$, the total spectral
density satisfies $\Jtot(E)=\Jinc(E)$
and has no interference zeros inside
the photonic band. The same phase that
eliminates distance-dependent delayed
feedback therefore also excludes
interference-induced BICs for two equal
coupling strengths.

This does not exclude bound states
outside the continuum. In a finite-band
waveguide, band-edge bound states can
remain present even at $\phi=\pi/2$.
Their energies depend on the full pole
equation rather than on the resonant
interference condition alone.
The positions of the off-band poles depend on the full self-energy, and their quantitative characterization in the USC regime requires retaining the local potential derived below.

\subsection{Effect of the local photonic potential}

We now retain the local potential $\Hloc$
in the projected Hamiltonian
of Eq.~\eqref{eq:H1full}. Its separable
structure allows the spectral problem
to be reduced to a scalar equation.

We consider an eigenstate
\begin{equation}
 \ket{\Psi_E}=u\ket e+\sum_kv_k\ket k,
 \qquad
 A=\sum_kf_k^*v_k.
 \label{eq:VlocAnsatz}
\end{equation}
The corresponding eigenvalue equations are
\begin{align}
 (E-\Dr)u&=2\Dr A,
 \label{eq:VlocAtomic}\\
 (E-\omega_k)v_k&=2\Dr f_k(u+A).
 \label{eq:VlocPhoton}
\end{align}
Introducing $I(z)=\sum_k|f_k|^2/(z-\omega_k)$, Eq.~\eqref{eq:VlocPhoton} gives $A=2\Dr I(E)(u+A)$, or $A=[2\Dr I(E)/(1-2\Dr I(E))]u$. Substitution into Eq.~\eqref{eq:VlocAtomic}
yields the pole equation for eigenstates
with a nonzero atomic component,
\begin{equation}
 E-\Dr-\frac{4\Dr^2I(E)}
 {1-2\Dr I(E)}=0.
 \label{eq:VlocPole}
\end{equation}

The self-energy obtained without the
local potential can be written as
$\Sigma(z)=4\Dr^2I(z)$.
Thus, neglecting $\Hloc$ corresponds
to replacing the denominator in
Eq.~\eqref{eq:VlocPole} by unity.

The local potential therefore modifies
the same scalar self-energy through
a rank-one correction. The complete
one-excitation pole equation can be
solved together with the self-consistent
polaron equation without diagonalizing
the full $N\times N$ photonic matrix.

Defining the self-energy with the local potential as
$\Sigma_{\mathrm{loc}}(z)=4\Dr^2 I(z)/[1-2\Dr I(z)]$,
its imaginary part inside the photonic band satisfies
\begin{equation}
 -2\operatorname{Im}\Sigma_{\mathrm{loc}}(E+\ii0^+)
 =\frac{\bar{\mathcal J}(E)}
 {|1-2\Dr I(E+\ii0^+)|^2}.
 \label{eq:VlocImag}
\end{equation}
Thus, for a regular point in the band, the local potential
modifies the on-shell linewidth and the real-energy pole
condition, but cannot create an interference-induced BIC
when $\bar{\mathcal J}(E)>0$. In particular, at
$\phi=\pi/2$ such a BIC remains excluded within the
projected polaron Hamiltonian, even when $\Hloc$ is retained.

\section{Laboratory-frame observables in the one-excitation sector}
\label{app:observables}
Let
\begin{equation}
 \ket{\psi(t)}_P=c(t)\ket e+\sum_k\chg{\beta_k}(t)\ket k
 \label{eq:obsAnsatz}
\end{equation}
be normalized in the polaron frame.
Here $c(t)$ and $\chg{\beta_k}(t)$ are Schr\"odinger-picture amplitudes for the effective polaron Hamiltonian, with the overall
$E_{\mathrm{GS}}$ phase omitted. They are related to the
interaction-picture amplitudes of Sec.~\ref{sec:decay} by
$c(t)=e^{-\ii\Dr t}c_e(t)$ and
$\chg{\beta_k}(t)=e^{-\ii\omega_k t}c_k(t)$.
Transforming the photon-number operator with Eq.~\eqref{eq:aTransformApp} gives
\begin{equation}
 n_k(t)=|f_k|^2+|\chg{\beta_k}(t)|^2
 -2\Re\left[f_kc(t)\chg{\beta_k}^*(t)\right].
 \label{eq:nkLab}
\end{equation}
The first term is the equilibrium cloud, the second is the propagating polaron-frame excitation, and the last term is their interference.

Within the same second-order, number-conserving truncation, expanding Eq.~\eqref{eq:sigmazexact} gives the laboratory atomic population
\begin{align}
 P_e(t)={}&P_{e,\mathrm{GS}}
 +\frac{\Dr}{\Delta}\Bigg[|c(t)|^2
 +2\Re\left(c(t)\sum_kf_k\chg{\beta_k}^*(t)\right)
 \nonumber\\
 &+2\sum_{k,p}f_kf_p^*\chg{\beta_k}^*(t)\chg{\beta_p}(t)\Bigg].
 \label{eq:PeLab}
\end{align}
These relations are useful for comparing the effective polaron dynamics with numerical simulations of the original Hamiltonian.

\bibliographystyle{apsrev4-1}
\bibliography{chiral_usc_refs}

\end{document}